\RequirePackage{fix-cm}
\documentclass[smallextended]{svjour3}       % onecolumn (second format)
\smartqed  % flush right qed marks, e.g. at end of proof
\usepackage{amssymb}
\usepackage{amsmath}
\usepackage{graphicx}
\journalname{\qquad  Acta Applicandae Mathematicae (2014) 132: \ pp. 347–-358}
\begin{document}

\title{Vesicle model with bending energy  revisited   %\thanks{Grants or other notes
}
\subtitle{}

\titlerunning{Vesicles with bending energy}        % if too long for running head

\author{Henri Gouin
}

\institute{ \at
             Aix-Marseille Universit\'e, CNRS, Centrale Marseille, M2P2 UMR 7340,\ \ 13451  Marseille, France \\
              \email{henri.gouin@univ-amu.fr; henri.gouin@yahoo.fr}             \\ }
\date{DOI: 10.1007/s10440-014-9907-y}

\maketitle

\begin{abstract}The equations governing
the conditions of mechanical equilibrium in fluid membranes subject to bending are revisited thanks to the principle of virtual work. The  note  proposes systematic tools to obtain the shape equation and the line condition
 instead of Christoffel symbols and the complex calculations they entail.
The method seems adequate to investigate all problems involving surface energies.
% Insert your abstract here. Include keywords, PACS and mathematical
% subject classification numbers as needed.
\keywords{Vesicles, surface energy, differential geometry of surfaces, shape equation.}
\PACS{45.20.dg,   68.03.Cd, 68.35.Gy,
02.30.Xx.}
\subclass{74K15, 76Z99, 92C37.}
\end{abstract}

\section{Introduction}
Lipid molecules dissolved in water spontaneously form bilayer membranes, with properties very similar to those of  biological membranes and vesicles \cite{Alberts}.
The knowledge of the mechanics of vesicles started
 more than thirty years ago when both experimental and theoretical studies of amphiphilic bilayers  engaged the attention of physicists and   the interest of mathematicians \cite{Lipowsky}.
In a viscous fluid,  vesicles are drops  a few tens of micrometers wide, bounded by   impermeable lipid membranes a few nanometers thick. The membranes are homogeneous down to molecular dimensions; consequently, it is possible to represent the vesicle as a two-dimensional smooth surface in   three-dimensional Euclidean space.
Depending on the cases, the bilayers  may be considered as liquid or solid. When liquid, the lipid molecules form  a two-dimensional lattice and   the membranes are described with an effective energy that does not penalize tangential displacements.
Their mechanical properties  permit a continuous mechanical description; such a  deformable object is characterized by a flexion-governed membrane rigidity resulting from the curvature energy.
The general theory accounts for surface strain, director extension, and director tilt associated with the misalignment of the surface normal. Galilean invariance is tantamount to the invariance of the energy under arbitrary two-dimensional orthogonal transformations and  regarded as a function of a symmetric two-dimensional tensor.  In this respect, the \emph{Helfrich  Hamiltonian}, which is quadratic in the curvature eigenvalues, provides a good description of lipid membranes and   associates bending with an energy penalty  \cite{Helfrich,Helfrich1,Seifert}. Equilibrium configurations of the membrane   satisfy a single normal 'shape' equation corresponding to the extrema of the Hamiltonian.
\newline
Fournier used a variational method to study fluid membranes \cite{Fournier}. Contrary wise, in this note,
we revisit the mechanical behaviour of vesicle membranes by using the principle
of virtual work \cite{Germain,Maugin,Gouin1} together with lemmas   from
the intrinsic differential geometry of surfaces; consequently, we do not need to use
coordinate lines and Christoffel symbols associated with the membrane metric are useless here.
For the Helfrich model, we established the equilibrium equation
and the boundary conditions - which even apply to compressible media and extensible membranes on cell surfaces - as well as the condition on a surface/vesicle contact line
  \cite{Steigmann1,Capovilla,Biscari,Napoli,Tu}.
For example,   in the case of bending energy, we obtain  the equilibrium equation of  the vesicle, the conditions on surfaces and line when the vesicle is in contact with a solid boundary; the condition on the vesicle membrane yields the 'shape' equation and a modified Young-Dupr\'e equation on the line.\\

\noindent The note is organized as follows:\\
In Section 2, we recall the  principle of virtual work. Section 3 introduces  universal geometrical tools associated  with the notion of virtual displacements introduced in Section 2.1 and the Stokes formula for volumes, surfaces and lines. The variations of tensorial quantities and differential forms on surfaces and lines are proposed in an intrinsic form, with no representation in  coordinate  lines.  Section 4 describes the vesicles' main background   and the consequence it has on the variation of the vesicle's energy. A conclusion focusing on the advantage of the intrinsic geometrical method ends the note.

\section{The principle of virtual work  applied to continuous media}
In continuum mechanics, the equilibrium of a medium can be  studied with either the Newton model of \emph{forces} or the Lagrange model of
 \emph{work of forces}. At equilibrium, a  minimization of the energy
 associated with a one-parameter  family corresponds with the zero value
of a linear functional of virtual displacements. The linear functional expressing
the  forces' work is related to the theory of distributions; a decomposition,
 associated with displacements considered as test functions whose supports are
compact manifolds,  uniquely determines a  separated form
respecting both  the test functions and their transverse derivatives  \cite{Schwartz}. Then, the equilibrium equation  and the boundary conditions
 are straightforwardly deduced from the principle of virtual work.

\subsection{\textbf{\emph{The notion of virtual displacement}}}

The position of a continuous medium is classically represented by a
transformation $\boldsymbol{\varphi }$ of a three-dimensional reference domain $D_0$ into the
physical set $D$. In order to describe  $\boldsymbol{\varphi }$ analytically, the
variables $\boldsymbol{X} = (X^{1},X^{2},X^{3})$ which single out individual
particles in $D_0$ correspond to   Lagrange's  variables; the variables $%
\boldsymbol{x}= (x^{1},x^{2},x^{3})$ in $D$ correspond to   Euler's   variables.  Transformation $\boldsymbol{\varphi}$ thus represents the position of a continuous medium,
\begin{equation*}
\boldsymbol{x}=\boldsymbol{\varphi} \left( \boldsymbol{X}\right) \text{ \quad or \quad }x^{i}=\varphi
^{i}(X^{1},X^{2},X^{3})\, , \ i \in \left\{1, 2, 3\right\},  \label{motion}
\end{equation*}
and possesses
inverse and continuous derivatives up to the second order except on singular
surfaces, curves or points.
To formulate the  principle of virtual work  in continuum
mechanics, we recall the notion of virtual displacements \cite{Serrin}:
 A one-parameter family of varied positions
possessing continuous partial derivatives up to the second order and
analytically expressed by the transformation
\begin{equation*}
\boldsymbol{x}=\boldsymbol{\Phi} \left( \boldsymbol{X},\eta \right)  \label{vitual
motion}
\end{equation*}
with $\eta \in O,$ where $O$ is an open real set containing $0$, and is such
that $\boldsymbol{\Phi }\left( \boldsymbol{X,}0\right) =\boldsymbol{\varphi }\left(
\boldsymbol{X}\right) $. The derivative, with respect to $\eta $ at $\eta =0$, is
noted  $\delta $ and is named
\textit{variation}  \cite{Serrin}\,.   In the physical
space, the virtual
displacement $\boldsymbol{\zeta}$ of a particle at $\boldsymbol{x}$ is such
that $\boldsymbol{\zeta}=\delta \boldsymbol{x}\ $ when  we assume $\delta\boldsymbol{X}=0$ and
  $\delta \eta =1$  at $\eta =0$; the virtual
displacement $\boldsymbol{\zeta}$ belongs to  $T_{\boldsymbol{x}}(D)$, a tangent vector bundle to $D$ at $\boldsymbol{x}$,
\begin{equation*}
{\boldsymbol{x}}\in D\  \longrightarrow \boldsymbol{\zeta}  =\boldsymbol{\psi (x)}%
\equiv \frac{\partial \boldsymbol{\Phi }}{\partial \eta }\left| _{\eta
=0}\right. \in T_{\boldsymbol{x}}(D).\label{virtualfield}
\end{equation*}

\subsection{\textbf{\emph{The background underlying the principle of virtual work}}}
The virtual work of forces $\delta\tau$   is a linear functional
value of the virtual displacement $\delta \boldsymbol{\varphi}$
determined by the variation of each particle and defined by
\begin{equation}
\delta \tau =<\Im ,\delta \boldsymbol{\varphi}>  \label{virtual work of
forces}
\end{equation}
where  $< \, , \, >$ denotes an inner product.
In Relation (\ref{virtual work of forces}),  $\delta \boldsymbol{\varphi}$ is submitted to  covector $\Im $ denoting all forces and stresses. Let us simply note that in
  case of motion, we must add  the inertial forces, corresponding to the
accelerations of masses,   to the volume forces, and eventually add the viscous stresses to the conservative stress tensor. The virtual displacements are naturally submitted to the constraints resulting from
constitutive equations  such as  mass conservation for compressible media. In this case, the
constraints are not necessarily expressed by Lagrange multipliers but are directly taken
into account by virtual displacements submitted to the variations of the constitutive equations. Conversely, when   geometrical assumptions are assumed, the Lagrange multipliers associated with  geometrical conditions constrain  the   virtual displacements, which in all cases are named  virtual displacements compatible with the constraints.\newline
The principle of virtual work is expressed in the form  :
\textit{{For all  virtual displacements compatible with the constraints, the virtual work of forces is null}.}
\\
If the distribution (\ref{virtual work of forces}) is in a separated form   \cite{Schwartz},   the principle of virtual work
 yields the equilibrium    (or motion) equation and the boundary conditions \cite{Gouin1}.

 \section{Intrinsic geometrical tools for the energy of surfaces and lines}
 We assume that   $D$  has a differential boundary $S$, except on its edge $C$. We respectively note $S_0$ and $C_0$ the images of $S$ and $C$ in   $D_0$; $D$ and $D_0$ are   Euclidian sets. The unit vector $\boldsymbol{n}$ and its image $\boldsymbol{n}_0$ are the oriented normal vectors to $S$ and
$S_0$; $c_m\equiv(R_{m})^{-1}$ is the mean curvature of $S$; the vector $\boldsymbol{t}$ is the oriented unit vector to $C$ and $\boldsymbol{n^{\prime}}= \boldsymbol{t}\times \boldsymbol{n}$ is the unit binormal vector \cite{Aris,Koba}.
The tensor $\boldsymbol{F} \equiv \partial \boldsymbol{x}/\partial \boldsymbol{X}$ denotes the Jacobian transformation of $\boldsymbol{\varphi}$; symbols\ $\rm{div},\,\rm{rot},\,  \rm{tr}$  and superscript $^T$ refer to the divergence, rotational,  trace operators and the transposition, respectively; $\textbf{1}$ denotes the identity tensor.

\begin{lemma}: we have the following relations
\begin{eqnarray}
\delta \det \boldsymbol{F} &=&\ \det \boldsymbol{F}\, \rm{div}\,\boldsymbol{\zeta}\,,  \label{Jacobi} \\
\delta \left( \boldsymbol{F}^{-1}\boldsymbol{n}\right) &=&-\boldsymbol{F}^{-1}\ \frac{\partial \boldsymbol{%
\zeta }}{\partial \boldsymbol{x}}\,\boldsymbol{n\ +\ }\boldsymbol{F}^{-1} \delta \boldsymbol{n}.  \label{InverseF}
\end{eqnarray}
\end{lemma}

\emph{Proof of} Rel. (\ref{Jacobi}): \newline

\noindent The   Jacobi identity written in the form\ \
$
\delta (\det \boldsymbol{F})=\det \boldsymbol{F}\,\mathtt{tr}\left( \boldsymbol{F}^{-1}\delta \boldsymbol{F}\right)
$
and
\begin{equation*}
\delta \boldsymbol{F}=\delta \left( \frac{\partial \boldsymbol{x}}{\partial \boldsymbol{X}}%
\right) =  \frac{\partial \boldsymbol{\zeta}}{\partial \boldsymbol{X}} ,
\end{equation*}%
imply,
\begin{equation*}
\mathtt{tr}\left( \boldsymbol{F}^{-1}\delta \boldsymbol{F}\right) =\mathtt{tr}\left( \frac{\partial
\boldsymbol{X}}{\partial \boldsymbol{x}}\,\frac{\partial \boldsymbol{\zeta}}{\partial
\boldsymbol{X}}\right) =\mathtt{tr}\left( \frac{\partial \boldsymbol{\zeta}}{%
\partial \boldsymbol{X}}\,\frac{\partial \boldsymbol{X}}{\partial \boldsymbol{x}}\right)
=\mathtt{tr}\left( \frac{\partial \boldsymbol{\zeta}}{\partial \boldsymbol{x}}%
\right) =\mathrm{div}\,{\boldsymbol{\zeta}}.
\end{equation*}
For an incompressible medium, $\rm{det}\, \boldsymbol{F} = 1$ and  $\boldsymbol{\zeta}$  verifies
\begin{equation}
\rm{div}\,\boldsymbol{\zeta} = 0 \label{incompressibleconstraint}.
\end{equation}

 \emph{Proof of} Rel. (\ref{InverseF}):  \newline
\begin{equation*}
\delta \left( \boldsymbol{F}^{-1}\boldsymbol{n}\right) =\delta \left( \boldsymbol{F}^{-1}\right) \boldsymbol{n}%
+\boldsymbol{F}^{-1}\delta \boldsymbol{n}
\end{equation*}
and
\begin{equation*}
\boldsymbol{F}^{-1}\,\boldsymbol{F}=\boldsymbol{1}\ \Longrightarrow \ \delta \left( \boldsymbol{F}^{-1}\right)
\boldsymbol{F}+\boldsymbol{F}^{-1}\,\delta \boldsymbol{F}=0,
\end{equation*}
imply
\begin{equation}
 \delta \left( \boldsymbol{F}^{-1}\right) =-\boldsymbol{F}^{-1}%
\frac{\partial \boldsymbol{\zeta}}{\partial \boldsymbol{X}}\boldsymbol{F}^{-1}= - \boldsymbol{F}^{-1}\frac{%
\partial \boldsymbol{\zeta}}{\partial \boldsymbol{x}}\ . \label{delta F-1}
\end{equation}

\begin{lemma}:
The variation of
$E=\protect\iint_{S}\protect\sigma\,ds$ is given by the relation
\begin{equation}
\delta E=\iint_{S}\left[ \delta \sigma-\left( \frac{2\sigma%
}{R_{m}}\,\boldsymbol{n}^{T}+\mathrm{{grad}^{T}\sigma\left( \boldsymbol{1}-\boldsymbol{nn}^{T}\right) }%
\right) \boldsymbol{\zeta}\right] ds + \int_C \sigma\, \boldsymbol{n^{\prime}}^T \boldsymbol{\zeta}~dl.  \label{varsurf0}
\end{equation}
where $\sigma$ is a scalar field defined on $S$ and $ds$, $dl$ are the surface and the line measures.
\end{lemma}

\emph{{Proof}} of Rel. (\ref{varsurf0}):\newline

\noindent  The normal vector field is locally extended in the
vicinity of $S$ by the relation ${\boldsymbol{n(x)}} = \mathrm{grad}\ d(\boldsymbol{x%
}), $ where $d$ is the distance of  point $\boldsymbol{x}$ to $S$; for any
vector field ${\boldsymbol{w}}$,
\begin{equation*}
\mathrm{rot} ({\boldsymbol{n}} \times {\boldsymbol{w}}) ={\boldsymbol{n }}\, \mathrm{div}
\, {\boldsymbol{w}}- {\boldsymbol{w}}\,\mathrm{div}\,{\boldsymbol{n }} + \frac {\partial
{\boldsymbol{n}}} {\partial {\boldsymbol{x}}}\, {\boldsymbol{w}}- \frac {\partial {%
\boldsymbol{w}}} {\partial {\boldsymbol{x}}}\, {\boldsymbol{n}}.
\end{equation*}
From $\,\displaystyle {\boldsymbol{n}}^T\frac {\partial {\boldsymbol{n}}} {\partial {%
\boldsymbol{x}}}= 0\,$ and $\,\mathrm{div}\,{\boldsymbol{n }} =\displaystyle -\frac {%
2} {R_m}$\,,  we deduce on $S$,
\begin{equation}
{\boldsymbol{n}^T} \mathrm{rot} ({\boldsymbol{n}} \times {\boldsymbol{w}}) = \mathrm{div}
\, {\boldsymbol{w}}+\frac {2} {R_m}\  {\boldsymbol{n}^T} {\boldsymbol{w}} - {\boldsymbol{n}^T%
} \frac {\partial {\boldsymbol{w}}} {\partial {\boldsymbol{x}}}\, {\boldsymbol{n}}.
\label{A0}
\end{equation}
Due to $\displaystyle\ E=\iint_S \sigma\ \det\, ({\boldsymbol{n}},d_1{\boldsymbol{%
x}},d_2{\boldsymbol{x}})\ $ where $\displaystyle\ d_1{\boldsymbol{x}} \ $ and $%
\displaystyle\ d_2{\boldsymbol{x}}\ $ are two coordinate lines of $S, $ we get,
\begin{equation*}
E =\iint_{S_0}\sigma\, \det  \boldsymbol{F}\ \hbox{det} (\boldsymbol{F}^{-1}{\boldsymbol{n}},d_1{%
\boldsymbol{X}},d_2{\boldsymbol{X}}) ,
\end{equation*}
with $d_1{\boldsymbol{x}}=\boldsymbol{F}\,d_1{\boldsymbol{X}}$,  $d_2{\boldsymbol{x}}=\boldsymbol{F}\,d_2{\boldsymbol{X}}$. Then,
\begin{equation*}
\begin{array}{cc}
\displaystyle \delta E =\iint_{S_0}\delta \sigma\ \det  \boldsymbol{F}\ \hbox{det}\,
(\boldsymbol{F}^{-1}{\boldsymbol{n}},d_1{\boldsymbol{X}},d_2{\boldsymbol{X}}) \\
\displaystyle +
\iint_{S_0}\sigma\, \delta\big(\det  \boldsymbol{F}\ \hbox{det}\, (\boldsymbol{F}^{-1}{\boldsymbol{n}%
},d_1{\boldsymbol{X}},d_2{\boldsymbol{X}})\big).
\end{array}%
\end{equation*}
Due to Lemma 1, $\,\displaystyle {\boldsymbol{n}}^T\frac {\partial {\boldsymbol{n}%
}} {\partial {\boldsymbol{x}}}= 0$ and to ${\boldsymbol{n}^T}\delta{\boldsymbol{n}}=0$,
\begin{equation*}
\begin{array}{cc}
\displaystyle \iint_{S_0}\sigma\, \delta\big(\det \boldsymbol{F}\ \hbox{det}\,
(\boldsymbol{F}^{-1}{\boldsymbol{n}},d_1{\boldsymbol{X}},d_2{\boldsymbol{X}})\big) = &  \\
\displaystyle \iint_S {\Big[}\sigma\ \mathrm{div}\, {\boldsymbol{\zeta}} \ \det ({%
\boldsymbol{n}},d_1{\boldsymbol{x}},d_2{\boldsymbol{x}}) + \sigma\, \det \left (%
\displaystyle  \delta \boldsymbol{n},d_1{\boldsymbol{x}},d_2{\boldsymbol{x}}\right ){\Big.} & \\
\displaystyle -\left.
\sigma \det \left (\displaystyle \frac {\partial{\boldsymbol{\zeta}}}{\partial {%
\boldsymbol{x}}}\,{\boldsymbol{n}},d_1{\boldsymbol{x}},d_2{\boldsymbol{x}} \right ) \right] ds= &  \\
\displaystyle \iint_S \left( \mathrm{div} (\sigma\,{\boldsymbol{\zeta}} )-(%
\mathrm{grad}^T \sigma) \, {\boldsymbol{\zeta}} -\sigma {\boldsymbol{n}}^T \frac {%
\partial {\boldsymbol{\zeta}}} {\partial {\boldsymbol{x}}} \, {\boldsymbol{n}} \right )
ds. &
\end{array}%
\end{equation*}
Relation (\ref{A0}) yields
\begin{equation*}
\displaystyle \mathrm{div}\, (\sigma\,{\boldsymbol{\zeta}}) + \frac {2\sigma} {%
R_m}\, {\boldsymbol{n}}^T {\boldsymbol{\zeta}} - {\boldsymbol{n}}^T \frac {\partial
(\sigma\, {\boldsymbol{\zeta}})} {\partial {\boldsymbol{x}}} \, {\boldsymbol{n}} = {%
\boldsymbol{n}}^T\, \mathrm{rot}\, (\sigma\,{\boldsymbol{n}}\times {\boldsymbol{\zeta}}
) . \qquad {\rm Then,}
\end{equation*}
\begin{equation}
\begin{array}{cc}
\displaystyle \iint_{S_0}\sigma\,\delta \big(\det  \boldsymbol{F}\,\hbox{det}\,
(\boldsymbol{F}^{-1}{\boldsymbol{n}},d_1{\boldsymbol{X}},d_2{\boldsymbol{X}})\big) = & \label{variationE} \\
\displaystyle\ \iint_{S} \left( - \frac {2\sigma} {R_m} \, {\boldsymbol{n}}^T+%
\mathrm{grad}^T\sigma\,({\boldsymbol{nn}^T-\boldsymbol{1}}) \right ) {\boldsymbol{\zeta}}
\, ds+\iint_S{\boldsymbol{n}}^T\ \mathrm{rot}\, (\sigma\,{\boldsymbol{n}}\times{%
\boldsymbol{\zeta}})\,ds, \notag &
\end{array}%
\end{equation}
where $\mathrm{grad}^{T}\sigma \,(\boldsymbol{1}-{{\boldsymbol{nn}}^{T}}) \equiv \mathrm{grad}_{tg}^{T}\,\sigma $ belongs to the cotangent plane to $S$ and
\begin{equation*}
\iint_S{\boldsymbol{n}}^T\ \mathrm{rot}\, (\sigma\,{\boldsymbol{n}}\times{%
\boldsymbol{\zeta}})\,ds =\int_C(\boldsymbol{t},\sigma\,{\boldsymbol{n}},\boldsymbol{\zeta})\,dl = \int_C \sigma\, \boldsymbol{n^{\prime}}^T \boldsymbol{\zeta}~dl.
\end{equation*}
Then, we   obtain  relation (\ref{varsurf0}).
%\paragraph{\textbf{Corollary} 1:}   If $S$ has no edge,
%\begin{equation}
%\delta \iint_{S}~ds=-\iint_{S}   \frac{2
%}{R_{m}}\,\boldsymbol{n}^{T}
 % \boldsymbol{\zeta}  ~ds.  \label{varsurf1}
%\end{equation}

\begin{lemma}: The variation of the internal energy is \newline
\begin{equation}
\delta \iiint_{D}\rho \ \alpha \,dv  = \iiint_{D}(\mathrm{grad}\,p)^{T}\,{\boldsymbol{\zeta}}\,dv-\iint_{S}{p\,%
\boldsymbol{n}}^{T}{\boldsymbol{\zeta}}\,ds.  \label{pressure}
\end{equation}
where $\rho $  is the  mass density, $\alpha (\rho )$  is the fluid specific energy, $p =\rho^2\displaystyle \frac{\partial\alpha}{\partial\rho}$ is the thermodynamical pressure \cite{Rocard} and  $dv$ is the measure of volume.
\end{lemma}

\emph{Proof} of Rel. (\ref{pressure}):\newline

\noindent $ \delta\iiint_{D}\rho \ \alpha
\,dv=\iiint_{D}\rho \ \delta \alpha \,dv$\ where\ $
 \delta \alpha = ({\partial \alpha }/{\partial \rho })\ \delta \rho $.
Due to  mass conservation,
\begin{equation}
\rho \ \mathrm{det}\,\boldsymbol{F}=\rho _{0}(\boldsymbol{X}),  \label{mass}
\end{equation}%
where $\rho _{0}$ is defined on $D_{0}$. The differentiation of Eq. (\ref%
{mass}) yields
\begin{equation*}
\delta \rho \,\mathrm{det}\,\boldsymbol{F}+\rho \,\delta (\mathrm{det}\,\boldsymbol{F})=0,\   \rm{and\ from\ Lemma\ 1,\ we \ get}
\end{equation*}%
\begin{equation*}
\delta \,\rho =-\rho \,\mathrm{div}\,{\boldsymbol{\zeta}}.
\end{equation*}%
Consequently, \ $\mathrm{div}\  (p\  {\boldsymbol{\zeta}}) = p \ \, \mathrm{div} \,  {\boldsymbol{\zeta}} + (\mathrm{grad}\,p)^{T}\,{\boldsymbol{\zeta}}$\, and  we deduce relation (\ref{pressure}).\newline

\noindent In the same way,
\begin{lemma}:
For  any scalar field  $p$ defined on $D$,
\begin{equation}
\delta \iiint_{D}p\ \mathrm{div}\,{\boldsymbol{\zeta}}\,  dv = -
\iiint_{D}(\mathrm{grad}\,p)^{T}\,{\boldsymbol{\zeta}}\,dv+\iint_{S}p\
\boldsymbol{n}^{T}{\boldsymbol{\zeta}}\,ds . \label{varsurf2}
\end{equation}
\end{lemma}
\begin{lemma}: The variation of the external unit vector normal to $S$ is
\begin{equation}
\delta \boldsymbol{n} = \left(\boldsymbol{n} \boldsymbol{n}^T -\textbf{1}\right) \left(\frac{\partial\boldsymbol{\zeta}}{\partial\boldsymbol{x}}\right)^T\boldsymbol{n} . \label{n variation}
\end{equation}
\end{lemma}

\emph{Proof} of Rel. (\ref{n variation}):\newline

\noindent The property $\Big\{\boldsymbol{n}^{T} d{%
\boldsymbol{x}} =0 \Longrightarrow \boldsymbol{n}^{T} \boldsymbol{F}\, d{\boldsymbol{X}} =0\Big\}$  implies that vector $\boldsymbol{F}^{T}\boldsymbol{n}$ is normal to $S_{0}$ and
consequently,    $\boldsymbol{n}_{0}^{T}\, \boldsymbol{n}_{0} = 1$ yields
\begin{equation*}
\boldsymbol{n}_{0} =\frac{\boldsymbol{F}^{T}\boldsymbol{n}}{\sqrt{(\boldsymbol{n}^{T}\boldsymbol{F}\boldsymbol{F}^{T}\boldsymbol{%
n})}} \,, \qquad \boldsymbol{n}  =\frac{{\boldsymbol{F}^{-1 T}} \boldsymbol{n}_0}{\sqrt{(\boldsymbol{n}^{T}_0\boldsymbol{F}^{-1} {\boldsymbol{F}^{-1 T}}  \boldsymbol{n_0})}}\ .
\end{equation*}
Then,  $\ \delta\boldsymbol{n}_0= 0$   on the reference surface  $S_0$ implies,
\begin{equation*}
\delta \boldsymbol{n}= \frac{\delta \boldsymbol{F}^{-1 T}\boldsymbol{n}_0}{\sqrt{\boldsymbol{n}_0^T \boldsymbol{F}^{-1}\boldsymbol{F}^{-1 T}\boldsymbol{n}_0}}
-\frac{1}{2}\,\boldsymbol{F}^{-1}\boldsymbol{n}_0\
\frac{\delta\left(\boldsymbol{n}_0^T \boldsymbol{F}^{-1}\boldsymbol{F}^{-1 T}\boldsymbol{n}_0\right)}
{\left(\boldsymbol{n}_0^T \boldsymbol{F}^{-1}\boldsymbol{F}^{-1 T}\boldsymbol{n}_0 \right)^{\frac{3}{2}}}\ .
\end{equation*}
From Eq. (\ref{delta F-1}), $\displaystyle \delta \boldsymbol{F}^{-1 T} = -\left(\frac{\partial\boldsymbol{\zeta}}{\partial\boldsymbol{x}}\right)^T\, \boldsymbol{F}^{-1 T}$ and consequently,
\begin{equation*}
\delta \boldsymbol{n} = -\left(\frac{\partial\boldsymbol{\zeta}}{\partial\boldsymbol{x}}\right)^T \boldsymbol{n} +\frac{1}{2}\,\boldsymbol{n}\left[\boldsymbol{n}^T \left(\frac{\partial\boldsymbol{\zeta}}{\partial\boldsymbol{x}}\right)\boldsymbol{n}+
\boldsymbol{n}^T \left(\frac{\partial\boldsymbol{\zeta}}{\partial\boldsymbol{x}}\right)^T\boldsymbol{n}\right] .
\end{equation*}
Then,\ \ $\displaystyle \boldsymbol{n}^T \left(\frac{\partial\boldsymbol{\zeta}}{\partial\boldsymbol{x}}\right)\boldsymbol{n}=
\boldsymbol{n}^T \left(\frac{\partial\boldsymbol{\zeta}}{\partial\boldsymbol{x}}\right)^T\boldsymbol{n}$\ \ implies  relation (\ref{n variation}).

\begin{lemma}: The variation of the mean curvature of $S$ is
\begin{equation}
\delta c_m = \frac{\partial c_m}{\partial\boldsymbol{x}}\,\boldsymbol{\zeta}+ \frac{1}{2}\,\Delta_{tg}(\boldsymbol{n}^T
\boldsymbol{\zeta}),\label{deltaH}
\end{equation}
where $\Delta_{tg}$ is the tangential Beltrami-Laplace operator on  surface $S$.
\end{lemma}

\emph{Proof} of Rel. (\ref{deltaH}):\newline

\noindent The variation of a derivative is given by
\begin{equation}
\delta\left(\frac{\partial\boldsymbol{n}}{\partial\boldsymbol{x}}\right) =  \frac{\partial\delta\boldsymbol{n}}{\partial\boldsymbol{x}}
- \frac{\partial\boldsymbol{n}}{\partial\boldsymbol{x}} \frac{\partial \boldsymbol{\zeta}}{\partial\boldsymbol{x}}. \label{varderi}
\end{equation}
From  $\displaystyle 2\,c_m = -\,\rm{div}\,  \boldsymbol{n}=-\rm{tr}  \left(\frac{\partial\boldsymbol{n}}{\partial\boldsymbol{x}}\right)$ and Eq. (\ref{varderi}) we get,
\begin{equation*}
2\,\delta c_m = - \rm{tr} \left(\frac{\partial\delta\boldsymbol{n}}{\partial\boldsymbol{x}}\right)
+ \rm{tr}\left(\frac{\partial\boldsymbol{n}}{\partial\boldsymbol{x}} \frac{\partial \boldsymbol{\zeta}}{\partial\boldsymbol{x}}\right)= - \rm{div}\, \delta\boldsymbol{n} + \rm{div} \left(\frac{\partial\boldsymbol{n}}{\partial\boldsymbol{x}}\, \boldsymbol{\zeta}\right)- 2\,\frac{\partial(\rm{div}\, \boldsymbol{n})}{\partial\boldsymbol{x}}
\,\boldsymbol{\zeta}.
\end{equation*}
But, $\displaystyle\frac{\partial(\rm{div}\, \boldsymbol{n})}{\partial\boldsymbol{x}}
= - 2\frac{\partial c_m}{\partial\boldsymbol{x}}$ and by using Eq. (\ref{n variation}) we get,
\begin{equation*}
- \rm{div}\, \delta\boldsymbol{n} + \rm{div} \left(\frac{\partial\boldsymbol{n}}{\partial\boldsymbol{x}}\, \boldsymbol{\zeta}\right)= \rm{div}\left[\left(\textbf{1}-\boldsymbol{n} \boldsymbol{n}^T \right)
\left(\frac{\partial\left(\boldsymbol{n}^T
\boldsymbol{\zeta}\right)}{\partial\boldsymbol{x}}\right)^T\right]= \rm{div\, grad_{tg}}(\boldsymbol{n}^T
\boldsymbol{\zeta}),
\end{equation*}
and from $ {\rm{div\, grad_{tg}}}(\boldsymbol{n}^T
\boldsymbol{\zeta})= \Delta_{tg}(\boldsymbol{n}^T
\boldsymbol{\zeta})$\
(\footnote{For all vector fields\ $\boldsymbol{x}\in D\rightarrow\boldsymbol{v(x)}$,\
$\displaystyle
\rm{div_{tg}} \boldsymbol{v} = \rm{div} \boldsymbol{v} - \boldsymbol{n}^T({\partial \boldsymbol{v}}/{  \partial\boldsymbol{x}})\,\boldsymbol{n} .
$
Then,
$ \displaystyle
\rm{div_{tg}} \boldsymbol{v} = \rm{div} \boldsymbol{v} - \rm{tr}\left(\boldsymbol{n}\boldsymbol{n}^T (\partial \boldsymbol{v} /{\partial\boldsymbol{x}})\right) .
$
But
$\displaystyle
\rm{div}\left(\boldsymbol{n}\boldsymbol{n}^T \boldsymbol{v}\right) = \rm{div}\left(\boldsymbol{n}\boldsymbol{n}^T \right)\boldsymbol{v}+\rm{tr}\left(\boldsymbol{n}\boldsymbol{n}^T ({\partial \boldsymbol{v}}/{  \partial\boldsymbol{x}})\right)
$
and
$\displaystyle
\rm{div}\left(\boldsymbol{n}\boldsymbol{n}^T \right)= \rm{div}\left(\boldsymbol{n} \right)\boldsymbol{n}^T+ \boldsymbol{n}^T \left({\partial \boldsymbol{n}}/{\partial\boldsymbol{x}}\right)^T .
$
From $\displaystyle \left({\partial \boldsymbol{n}}/{\partial\boldsymbol{x}}\right) \boldsymbol{n} = 0$ and $\displaystyle \rm{div} \boldsymbol{n} = - {2}/{\it{R_m}}$, we get
$\displaystyle
\rm{div_{tg}} \boldsymbol{v} = \rm{div}\left((\textbf{1}-\boldsymbol{n}\boldsymbol{n}^T)\,\boldsymbol{v}\right) -  {2\, \boldsymbol{n}^T \boldsymbol{v}}/{\it{R_m}}
$.
From $\rm{div_{tg}} \boldsymbol{v}_{tg} = \rm{div} \boldsymbol{v}_{tg}$ where $\boldsymbol{v}_{tg} =
(\textbf{1}-\boldsymbol{n}\boldsymbol{n}^T)\boldsymbol{v}$, we get
$ \rm{div}\,\rm{grad_{tg}}(\boldsymbol{n}^T
\boldsymbol{\zeta})= \rm{div_{tg}}\,\rm{grad_{tg}}\,(\boldsymbol{n}^T\boldsymbol{\zeta}) = \Delta_{tg}(\boldsymbol{n}^T\boldsymbol{\zeta})
$.}), we deduce  relation (\ref{deltaH}).

\section{\textbf{Description of a vesicle membrane in contact with a solid surface}}

\subsection{\bf \emph{Membranes' bending energy}}
Vesicles consist in a three-dimensional domain bounded by a liquid bilayer.  Vesicle interfaces are represented by material surfaces endowed with
a bending surface energy. In our representation, a vesicle fills  set $D$ and lies on the surface of a solid; the
vesicle is also surrounded by a fluid (see Fig. \ref%
{fig1}). All the interfaces between vesicle,
solid and liquid are assumed to be regular.
\begin{figure}[h]
\begin{center}
\includegraphics[width=10cm]{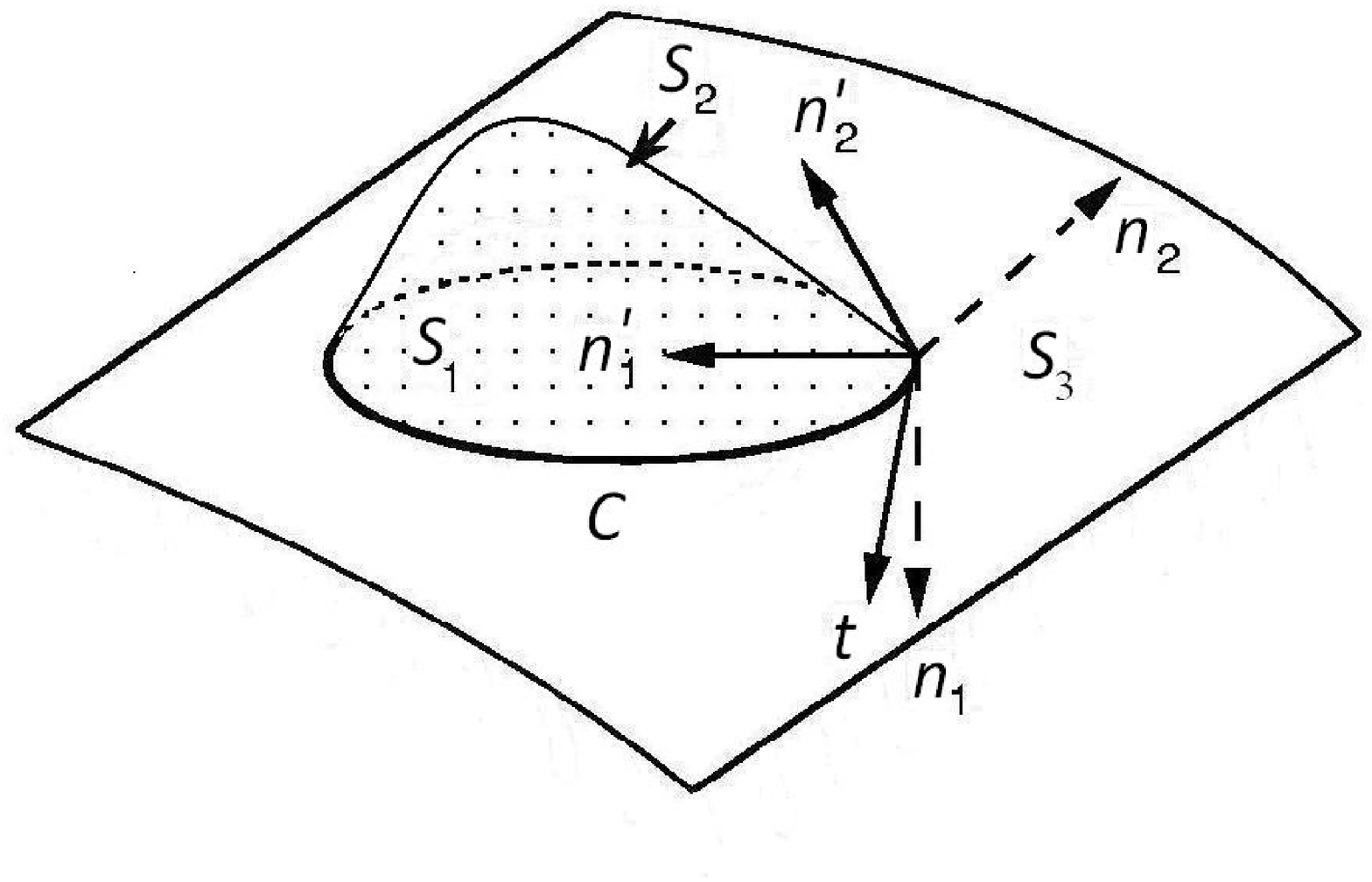}
\end{center}
\caption{ A drop-shaped vesicle
lies on a solid surface. The vesicle is bordered by a fluid (liquid) and a solid; $S{%
_1}$ is the boundary between liquid/solid; $S_{2}$ is the interface between vesicle/fluid  (membrane of the vesicle); $S_{3}$ is the
boundary between fluid/solid; $\boldsymbol{n}_1$ and $\boldsymbol{n}_2$ are the unit normal vectors to $%
S_{1} $ and $S_{2}$, external to the domain of the vesicle; contact line $C$  is shared by $S_{1}$ and $%
S_{2}$ and $\boldsymbol{t}$ is the unit tangent vector to $C$ relative to $%
\boldsymbol{n}_1$; $\boldsymbol{n^{\prime}}_1 = \boldsymbol{n}_{1}\times \boldsymbol{t}$ and $\boldsymbol{n^{\prime}}_2= \boldsymbol{t}\times \boldsymbol{n}_{2}$ are the
binormals to $C$ relative to $S_{1}$ and $S_{2}$, respectively. }
\label{fig1}
\end{figure}
We note  $\sigma_{1}$, $\sigma_{2}$ and $\sigma_{3}$ the values of the
surface energies of $S_{1}$, $S_{2}$ and $S _{3}$, respectively (\footnote{Our aim is not to consider the thermodynamics of interfaces. Consequently $\sigma_i$, $i\in \{1,2,3\}$ are not taken into account as a function  of variables like  temperature or   entropy.}).  The vesicle is submitted to
a volume force $\rho\, \boldsymbol{f}$; $S$ = $S_1 \cup S_2$ is the boundary of $D$; the external surface force
on $D$ is modelized with two  vector fields $\boldsymbol{T}_{1}$ on
the solid surface $S _{1} $ and $\boldsymbol{T}_{2}$ on the free vesicle surface $S _{2}$. In our calculus, the line tension on $C$ is assumed to  be null.
Vector $\boldsymbol{n}$   stands  for the normal to $S$ external to $D$.
\\
To obtain the equilibrium equation   and the boundary conditions, it is necessary to propose a constitutive behaviour for the membrane's energy density $\sigma_{2}$. As proved in the literature, the surface energy density on $S_2$ must be a function of the curvature tensor along the surface. To be   intrinsic, $\sigma_{2}$ must be a function of the two curvature tensor invariants.  If we note  $c_1= 1/R_1$ and $c_2=1/R_2$ the eigenvalues of the curvature tensor,
 the mean curvature and the Gauss curvature of $S_2$ are respectively noted,
\begin{equation*}
H = \frac{c_1+c_2}{2}=\frac{1}{R_m}, \qquad K =  {c_1\, c_2} =\frac{1}{R_1\,R_2}.
\end{equation*}
The external normal ${\boldsymbol{n_2}}$ to $S_2$ can be locally extended in the
vicinity of $S_2$ by the relation ${\boldsymbol{n_2(x)}} = \mathrm{grad}\ d_2(\boldsymbol{x%
}), $ where $d_2$ is the distance of  point $\boldsymbol{x}$ to $S_2$ (see Section 3, Lemma 2).
Then,
\begin{equation}
2H =-\, {\rm{div}}\,  {\boldsymbol{n}}_2\equiv -\, {\rm{tr}}\left(\frac{\partial\boldsymbol{n}_2}{\partial\boldsymbol{x}}\right),\quad \frac{\partial\boldsymbol{n}_2}{\partial\boldsymbol{x}}= \left(\frac{\partial\boldsymbol{n}_2}{\partial\boldsymbol{x}}\right)^T, \quad
{\boldsymbol{n}_2}^T\frac {\partial {\boldsymbol{n}_2}} {\partial {%
\boldsymbol{x}}}= 0, \label{H}
\end{equation}
\begin{equation*}
2K = \left[\rm{tr}\left(\frac{\partial\boldsymbol{n}_2}{\partial\boldsymbol{x}}\right)\right]^2-
\,\rm{tr}\left(\frac{\partial\boldsymbol{n}_2}{\partial\boldsymbol{x}}\right)^2.\label{K}
\end{equation*}
The surface's energy density $\sigma_{2}$ is assumed to be a regular function of $H$ and $K$, but in the Helfrich model, the vesicle's surface energy is linear in $K$.  The Gauss-Bonnet theorem ensures that the integration term corresponding to $K$ is constant
 for closed surfaces, otherwise, the
geodesic curvature of the boundary plays a role; this means, as established in \cite{Fournier},
that the Gaussian curvature affects the boundary line actions. Nonetheless, experimental and theoretical studies have shown that the energy mainly stems from the bending \cite{Lipowsky}; consequently, in the Helfrich model,  the surface energy density on $S_2$ is taken as a form without term in $K$:
\begin{equation}
\sigma_{2} = \sigma_{o}+\frac{\kappa}{2}\left[2H-c_o\right]^2, \label{Helfrich}
\end{equation}
where $\kappa$ is the bending rigidity, $c_o$ is the spontaneous curvature and $\sigma_o$ is the superficial energy  of capillarity.
The main interest being  the membrane's bending energy and its behaviour, we assume that the values of $\sigma_o$, $\sigma_{1}$ and $\sigma_{3}$ are  constant. This special case can be easily extended, as  done in \cite{Gouin2} for an other problem of capillarity.
\begin{lemma}:  The variation of the bending energy $E_2 = \iint_{S_2} \sigma_2\, ds$ of a membrane is \ \ $\delta E_2  =$
 \begin{equation}
\iint_{S_2}\left[\frac{d\sigma_2}{dn_2} - 2 H\,\sigma_2 +
 \frac{1}{2}\, \Delta_{tg}\left(\frac{\partial\sigma_2}{\partial H}\right)\right] \boldsymbol{n}_2^T
\boldsymbol{\zeta}\, ds  + \int_C \boldsymbol{n}_2^{\prime T} \left(\sigma_2\,\boldsymbol{\zeta}+\boldsymbol{w}\right)  dl \label{deltaE2},
\end{equation}
where $\displaystyle
 \boldsymbol{w}= \frac{1}{2}\left[\frac{\partial\sigma_2}{\partial H}\  {\rm{grad_{tg}}} (\boldsymbol{n}_2^T
\boldsymbol{\zeta})-\boldsymbol{n}_2^T
\boldsymbol{\zeta}\ {\rm{grad_{tg}}}\left(\frac{\partial\sigma_2}{\partial H}\right)\right]$
and
$\displaystyle\frac{d}{dn_2}\equiv \left(\frac{\partial  }{\partial \boldsymbol{x}}\right)\, \boldsymbol{n}_2$ is the normal derivative in the direction $\boldsymbol{n}_2$.
\end{lemma}

\emph{Proof} of Rel. (\ref{deltaE2}):\newline

\noindent From Eq. (\ref{varsurf0}), $\sigma_2 = \sigma_2(H)$ and Eq. (\ref{deltaH}), we obtain,
\begin{eqnarray*}
\delta E_2 = \iint_{S_2}\left[\frac{\partial\sigma_2}{\partial H}\,\frac{\partial H}{\partial\boldsymbol{x}}\,\boldsymbol{\zeta}+
\frac{1}{2}\,\frac{\partial\sigma_2}{\partial H}\,\rm{div\, grad_{tg}}(\boldsymbol{n}_2^T
\boldsymbol{\zeta})\right.\\ \Big. -2\,\sigma_2 H \boldsymbol{n}_2^T
\boldsymbol{\zeta}- \rm{grad}\,\sigma_2 \left(\textbf{1}-\boldsymbol{n}_2 \boldsymbol{n}_2^T \right)\boldsymbol{\zeta}\Big] ds +\int_C \sigma_2\, \boldsymbol{n}_2^{\prime T} \boldsymbol{\zeta}\, dl
\end{eqnarray*}
But, $\displaystyle \rm{grad}\,\sigma_2 = \frac{\partial\sigma_2}{\partial \emph{H}}\,\frac{\partial \emph{H}}{\partial\boldsymbol{x}}$; then,
\begin{equation*}
\delta E_2 = \iint_{S_2}\left[\frac{1}{2}\,\frac{\partial\sigma_2}{\partial \emph{H}}\ \rm{div}\, grad_{tg} (\boldsymbol{n}_2^T
\boldsymbol{\zeta})+ \left(\frac{\emph{d}\sigma_2}{\emph{d}\emph{n}_2}- 2\,\emph{H}\, \sigma_2\right)\boldsymbol{n}_2^T
\boldsymbol{\zeta} \right] ds +\int_C \sigma_2 \boldsymbol{n}_2^{\prime T} \boldsymbol{\zeta}\, dl,
\end{equation*}
and from
 \begin{equation*}
\quad \frac{1}{2}\,\frac{\partial\sigma_2}{\partial H}\  {\rm{div\,grad_{tg}}} (\boldsymbol{n}_2^T
\boldsymbol{\zeta}) =  \frac{1}{2}\, \Delta_{tg}\left(\frac{\partial\sigma_2}{\partial H}\right)\,\boldsymbol{n}_2^T
\boldsymbol{\zeta}+ \rm{div} \, \boldsymbol{w},
\end{equation*}
we deduce relation (\ref{deltaE2}).

\subsection{\textbf{\emph{Expression of the virtual work of forces}}}

For the Helfrich model, the total energy of the vesicle writes
\begin{equation*}
\Xi=\iiint_{D}\rho \ \alpha (\rho )\ dv+ \iint_{S_1}\sigma_1 \ ds + \iint_{S_2}\sigma_2 \ ds .
\end{equation*}%
The virtual work of volume force $\rho\, \boldsymbol{f}$  defined on $D$ writes
$
\iiint_{D}\rho \, \boldsymbol{f}^{T}\boldsymbol{\zeta}~dv
$.
 The virtual work of surface   force $\boldsymbol{T}$ exerted  on $S$
  is
$
\iint_{S} \ \boldsymbol{T}^{T}\boldsymbol{\zeta}~ds.
$
Due to Eq. (\ref{varsurf0}), $-\int_C \sigma_{3}\,\boldsymbol{n^{\prime}}_1^T \boldsymbol{\zeta}~dl$ corresponds to the action of $S_3$ on edge $C$.
Finally, the virtual work of forces writes
\begin{equation}
\delta\tau=-\delta \Xi +\iiint_{D}\rho \, \boldsymbol{f}^{T}\boldsymbol{\zeta}%
~dv+\iint_{S} \, \boldsymbol{T}^{T}\boldsymbol{\zeta }~ds-\int_C \sigma_{3}\,\boldsymbol{n^{\prime}}_1^T \boldsymbol{\zeta}~dl.  \label{Energy1}
\end{equation}
From Eqs.  (\ref{varsurf0}), (\ref{pressure}),  (\ref{deltaE2})  and (\ref{Energy1}), we obtain
\begin{eqnarray}
\delta \tau  &=&\iiint_{D} \left(\rho\, \boldsymbol{f}^T - {\rm{grad}}^{T}p\right)\boldsymbol{\zeta}~dv  \notag\\
&&+\iint_{S_1} \left[\left(p+\frac{2\, \sigma_{1}}{R_{m_1}}\right)\boldsymbol{n}_1^T+\boldsymbol{T}_1^T\right]\boldsymbol{\zeta}~\rm{ds} \notag\\
&&+\iint_{S_2} \left\{\left[p-\frac{d\sigma_{2}}{dn_2} + 2H\, \sigma_{2}- \frac{1}{2}\,\Delta_{tg}\left(\frac{\partial\sigma_2}{\partial H}\right)\right] \boldsymbol{n}_2^T\, +\boldsymbol{T}_2^T\right\} \boldsymbol{\zeta}~ds \notag \\
&&+\int_C \left[\left((\sigma_{1}-\sigma_{3})\,\boldsymbol{n^{\prime}}_1^T -\sigma_{2}\, \boldsymbol{n^{\prime}}_2^T \right) \boldsymbol{\zeta} - \boldsymbol{n^{\prime}}_2^T  \, \boldsymbol{w} \right]~dl \label{Laplace},
\end{eqnarray}%
where $2/R_{m_1}$ is the mean curvature of $S_1$, and $\boldsymbol{T}_1$ and  $\boldsymbol{T}_2$   correspond to the surface forces exerted  on $S_1$ and $S_2$, respectively.

\section{\textbf{Equations  governing  equilibrium and boundary conditions}}

The fundamental lemma of variational calculus applied to each integral of Eq. (\ref{Laplace}) yields  the equilibrium equation
 associated with  domain $D$, the conditions  on  surfaces  $S_1$ and $S_2$ and the   condition on contact line $C$.

\paragraph{\bf \emph{Equilibrium equation in} $D$}
\begin{equation*}
  {\rm{grad}}\,
p =    \rho \,\boldsymbol{f}.  \label{EE}
\end{equation*}
This  is the classical condition for equilibrium.
\paragraph{\bf \emph{Condition  on surface} $S_1$}
\begin{equation*}
\left(p+\frac{2\, \sigma_{1}}{R_{m_1}}\right)\boldsymbol{n}_1 +\boldsymbol{T}_1 = 0.
\end{equation*}
Then $T_1 = -p_1\, \boldsymbol{n}_1$ is  a normal stress vector to  surface $S_1$ and we obtain the classical Laplace condition,
\begin{equation*}
 p_1 -p =\frac{2\, \sigma_{1}}{R_{m_1}}     .
\end{equation*}
\paragraph{\bf \emph{Condition  on membrane surface} $S_2$}
\begin{equation}
\left[\,p-\frac{d\sigma_{2}}{dn_2} + 2H\, \sigma_{2}- \frac{1}{2}\,\Delta_{tg}\left(\frac{\partial\sigma_2}{\partial H}\right)\right] \boldsymbol{n}_2\, +\boldsymbol{T}_2 = 0 \label{membrane1}.
\end{equation}
Then, the   stress vector must be normal to $S_2$. In fact $\boldsymbol{T}_2 = - p_2\,\boldsymbol{n}_2$ corresponds to the action of the external fluid on the membrane.
From $\sigma_2 =\sigma_2 (H)$, and
taking   Eq. (\ref{H}) into account,
\begin{equation*}
\frac{d\sigma_2}{dn_2}= \frac{\partial\sigma_2}{\partial H}\, \frac{dH}{dn_2} \quad {\rm with}\quad 2\,\frac{\partial H}{\partial \boldsymbol{x}}= - \, {\rm{div}}  \left(\frac{\partial  \boldsymbol{n}_2}{\partial \boldsymbol{x}}\right)
\end{equation*}
and
\begin{equation*}
2\, \frac{dH}{dn_2}\equiv 2\,\frac{\partial H}{\partial \boldsymbol{x}}\, \boldsymbol{n}_2 =
- \,{\rm{div}}\, \left(\frac{\partial  \boldsymbol{n}_2}{\partial \boldsymbol{x}}\right)\boldsymbol{n}_2= -\, {\rm{div}}\, \left(\frac{\partial  \boldsymbol{n}_2}{\partial \boldsymbol{x}}\,\boldsymbol{n}_2\right)+ {\rm{tr}}\left(\frac{\partial  \boldsymbol{n}_2}{\partial \boldsymbol{x}}\right)^2.
\end{equation*}
Due to the fact that $\displaystyle \frac{\partial  \boldsymbol{n}_2}{\partial \boldsymbol{x}}\, \boldsymbol{n}_2= 0$, from Eq. (\ref{H}) we get  $\displaystyle \frac{d H}{dn_2} = 2\, H^2 - K$ and
\begin{equation*}
\frac{d  \sigma_2}{dn_2} = (2\, H^2 - K)\frac{\partial  \sigma_2}{\partial H}.
\end{equation*}
Finally,  Eq. (\ref{membrane1}) yields,
\begin{equation}
 \,p-p_2- (2\, H^2 - K)\frac{\partial  \sigma_2}{\partial H}  + 2H\, \sigma_{2}- \frac{1}{2}\,\Delta_{tg}\left(\frac{\partial\sigma_2}{\partial H}\right)  = 0 \label{membrane1bis}.
\end{equation}
In the case of the Helfrich model (\ref{Helfrich}),   we obtain from Eq. (\ref{membrane1bis}) the 'shape' equation (\footnote{Let us note that Helfrich \emph{et al} \cite{Helfrich,Helfrich1,Seifert} consider the vesicle as incompressible
and the virtual displacement verifies
$
\rm{div}\, \boldsymbol{\zeta} = 0
$
(see  Eq. (\ref{incompressibleconstraint})). They assume  that the lipid bilayer $S$ has a total constant area $S_0$ and introduce the   constraint
$
\iint_{S} ds =S_0.
$
Then,   the   virtual work is expressed as
\begin{equation*}
\delta \tau =\iiint_{D}\rho \ \boldsymbol{f}^{T}\boldsymbol{\zeta}%
~dv+  \iint_{S} \ \boldsymbol{T}^{T}\boldsymbol{\zeta}~ds -\,\delta\iint_S \sigma~ds \, + \lambda_0 \, \delta\iint_S ds + \,  \delta   \iiint_D p\ \rm{div}\ \boldsymbol{\zeta}\ dv ,\label{virtualwork}
\end{equation*}
where the scalar  $\lambda_0$ is a constant  Lagrange multiplier  and $p$\,   is a distributed  Lagrange multiplier. Due to Eq. (\ref{varsurf2}), the 'shape' equation that they deduced is identical to Eq. (\ref{membrane2}).}):
\begin{equation}
  p-p_2 + \kappa \left(2\,H-c_o\right)\left(2K-2H^2-c_oH\right)+2\,H\,\sigma_o -2\kappa \Delta_{tg}H = 0 \label{membrane2}.
\end{equation}

\paragraph{\bf \emph{Condition  on  line} $C$}.\\

 \noindent To get the line condition, we must consider a virtual displacement  tangent to the fixed surface $S_1$ and consequently
$
\boldsymbol{\zeta}=\alpha\, \boldsymbol{t}+\beta\, \boldsymbol{t}\times \boldsymbol{n}_{1},
$
where $\alpha $\ and $\beta $\ are two scalar fields defined on $C$. From the
last integral of Eq. (\ref{Laplace}), we get immediately: \
For any scalar field $\boldsymbol{x}\in  C \longrightarrow \beta \left(
\boldsymbol{x}\right) \in \Re ,$
\begin{equation*}
\int_{C} \big[ \big(\left(\sigma_1-\sigma_3\right)\boldsymbol{n^{\prime}}_1^T-\sigma_2\boldsymbol{n^{\prime}}_2^T \big )\boldsymbol{%
\zeta }- \boldsymbol{n^{\prime}}_2^T\boldsymbol{w}\big ]~dl=0,
\end{equation*}%
Due to the fact that\
$
\boldsymbol{n}^T_2\,\boldsymbol{\zeta}=\beta\,\boldsymbol{n}^T_2\,(\boldsymbol{t}\times \boldsymbol{n}_{1}) = \beta\, \boldsymbol{t}^T\,(\boldsymbol{n}_1\times \boldsymbol{n}_{2})= \beta\, {\rm sin}\,\theta ,
$                                                      where  $\theta = (\,\boldsymbol{n}_1,\boldsymbol{n}_2)$  is the Young angle, and the term\ $\beta\, {\rm sin}\,\theta$\ is uniquely function of  arc length $l$, we get $\boldsymbol{n^{\prime}}^T_2 {\rm grad_{tg}}(\beta\, {\rm sin}\,\theta) = 0$. Consequently, from Lemma 7,
\begin{equation*}
\boldsymbol{n^{\prime}}_2^T\boldsymbol{w}= -\frac{1}{2} \left(\boldsymbol{n}_2^T
\boldsymbol{\zeta}\right)\ \boldsymbol{n^{\prime}}^T_2\, {\rm grad_{tg}} \left(\frac{\partial\sigma_2}{\partial \emph{H}} \right)= -\frac{1}{2}\,\beta\, {\rm sin}\,\theta\, \sigma_2''(\emph{H})\,\frac{\emph{dH}}{\,\emph{d}{\emph{n}^{\prime}}_2},
\end{equation*}
where $\displaystyle \frac{dH}{\,d{n^{\prime}}_2}$ is the value of  the derivative of $H$   along the line orthogonal to $C$ on $S_2$.
Consequently, $\forall\,\{ l \longrightarrow \beta(l)\in \Re\} $,
\begin{equation*}
\int_{C} \beta\left\{\left(\sigma_1-\sigma_3\right)\boldsymbol{n^{\prime}}^T_1(\boldsymbol{t}\times \boldsymbol{n}_{1})-\sigma_2\boldsymbol{n^{\prime}}^T_2(\boldsymbol{t}\times \boldsymbol{n}_{1}) + \frac{\sigma_2''(H)}{2}\,  \frac{dH}{\,d{n^{\prime}}_2}\,{\rm sin}\, \theta  \right\}~dl=0.
\end{equation*}
Then, $$
 \int_{C} -\beta\left\{\left(\sigma_1-\sigma_3\right)+\sigma_2\boldsymbol{n}^T_2  \boldsymbol{n}_{1} -\frac{\sigma_2''(H)}{2}\,  \frac{dH}{\,d{n^{\prime}}_2}\,{\rm sin}\, \theta \right\}~dl=0,
$$
and we  obtain the line condition
\begin{equation}
\left(\sigma_1-\sigma_3\right)+\sigma_2\,{\rm cos}\, \theta - \frac{\sigma_2''(H)}{2}\,  \frac{dH}{\,d{n^{\prime}}_2}\,{\rm sin}\, \theta = 0
\label{Young1}.
\end{equation}
In the Helfrich model (\ref{Helfrich})    the condition (\ref{Young1}) yields
\begin{equation}
\left(\sigma_1-\sigma_3\right)+\sigma_2 \,{\rm cos}\, \theta -2\, \kappa\,    \frac{dH}{\,d{n^{\prime}}_2}\, {\rm sin}\,  \theta = 0 \label{Young2}.
\end{equation}
Condition (\ref{Young2}) replaces  the classical Young-Dupr\'e condition by taking  additive term $\displaystyle
- 2\,\kappa\,    \frac{dH}{\,d{n^{\prime}}_2}\, {\rm sin}\,  \theta $\ \,into account.
\section{\ Conclusion and remarks}
In this  note, we propose simple systematic  tools  coming from   surface geometry and  from the principle of virtual work  to obtain the boundary conditions on surfaces and lines for three-dimensional domains where the surfaces are endowed with  surface energy densities. The tools are based on the Helfrich model with bending energy  which is usually proposed to study the mechanics of  biological membranes. The model does not take  line energy  into account, but the calculations will be similar to obtain the conditions at the boundaries of three-dimensional domains.
Relation (\ref{varsurf0}) is the key point of the model and highlights the extreme importance of knowing the
variation of $\delta \sigma$ and consequently the  behaviour of the surface energy $\sigma$. For example,  in \cite{Gouin2} we obtained a case where the capillary surface energy   depended on the composition of the surface layer. The obtained results  do not need to assume  vesicle incompressibility and  constant area of the membrane. We notice that  the calculations proposed in the literature use the Christoffel symbols associated with coordinate curves on the surfaces,  but the Christoffel symbols  do not appear in the resulting expressions of the boundary conditions. This is an important reason for the straightforwardness of our method.


\begin{thebibliography}{99}

\bibitem{Alberts}Alberts,  B., Johnson, A., Lewis, J., Raff, M., Roberts, K.,  Walter, P.: Molecular biology of the cell, 4th edn.  Garland Science, New York (2002)

\bibitem{Lipowsky} Lipowsky, R., Sackmann, E. (eds.):   Structure and dynamics of membranes, Handbook of Biological Physics. Vol. 1A and Vol. 1B,  Elsevier, Amsterdam (1995)








\bibitem{Helfrich} Helfrich,  W.:  Elastic properties of lipid bilayers:
theory and possible experiments, Z. Naturforsch. C \textbf{28}, 693--703 (1973)

\bibitem{Helfrich1} Ou-Yang Zhong Can,  Helfrich, W.:  Bending energy of vesicle membranes: General expressions for the first, second, and third variation of the shape energy and applications to spheres and cylinders, Physical Review A  \textbf{39}, 5280--5288 (1989)

\bibitem{Seifert} Seifert, U.:  Configurations of fluid membranes and
vesicles, Adv. Phys. \textbf{46}, 13--137  (1997)

\bibitem{Fournier} Fournier J.B.: On the stress and torque tensors in fluid membranes, Soft Matter   \textbf{3}, 883–-888 (2007)

\bibitem{Germain} Germain, P.:  The method of the virtual power in
continuum mechanics - Part 2 : microstructure, SIAM  J. Appl. Math.
\textbf{25}, 556-575 (1973)

\bibitem{Maugin} Daher, N., Maugin, G. A.:  The method of virtual power in continuum
mechanics: application to media presenting singular surfaces and interfaces,
Acta Mechanica.  \textbf{60}, 217-240 (1986)

\bibitem{Gouin1} Gouin, H.:  The d'Alembert-Lagrange principle for
gradient theories and boundary conditions, in: Ruggeri, T., Sammartino, M.
(eds.), Asymptotic Methods in Nonlinear Wave Phenomena, p.p. 79-95, World Scientific,
 Singapore  (2007) \& arXiv:0801.2098



\bibitem{Steigmann1} Steigmann, D.J., Li,  D.:   Energy minimizing states of capillary systems with bulk, surface and line phases,  IMA J. Appl. Math. \textbf{55}, 1-17  (1995)

\bibitem{Capovilla} Capovilla, R., Guven, J.:  Stresses in lipid membranes, J. Phys. A: Math. Gen. \textbf{35},  6233–6247 (2002)

\bibitem{Biscari}   Biscari P., Canavese S.M., Napoli G.: Impermeability effects in three-dimensional vesicles, J. Phys. A: Math. Gen. \textbf{37}, 6859--6874 (2004)

\bibitem{Napoli}  Napoli, G., Vergori, L.:  Equilibrium of nematics vesicles, J.  Phys. A: Math.  Theo. \textbf{43}, 445207 (2010)

\bibitem{Tu} Tu,  Z. C.:  Geometry of membranes, J. Geom. Symm. Phys. \textbf{24}, 45-75 (2011)


\bibitem{Schwartz}  Schwartz, L.:  Th\'{e}orie des distributions. Ch.  3,
Hermann, Paris (1966)

\bibitem{Serrin} Serrin, J.:  Mathematical principles of classical fluid
mechanics, in Encyclopedia of Physics VIII/1,   S. Fl\"ugge (ed.), p.p. 125-263,
Springer, Berlin (1960)

\bibitem{Aris} Aris, R.:  Vectors, tensors, and the basic equations of
fluid mechanics, Dover Publications, New York  (1989)

\bibitem{Koba} Kobayashi, S., Nomizu, K.:  Foundations of differential
geometry, vol. 1, Interscience Publ., New York (1963)


\bibitem{Rocard} Rocard, Y.: Thermodynamique, Masson, Paris (1952)

\bibitem{Gouin2} Gouin,  H.:  Interfaces endowed with non-constant surface energies revisited with the d'Alembert-Lagrange principle, Mathematics and Mechanics of Complex Systems, \textbf{2},  (1) 23-43 (2014)    \& arXiv:1311.1140







\end{thebibliography}
\end{document}